\documentclass[%
reprint,showpacs,
superscriptaddress,
 amsmath,amssymb,
 aps,
]{revtex4-1}

\usepackage{multirow}
\usepackage{graphicx}
\usepackage{booktabs}
\usepackage{dcolumn}
\usepackage{amsmath}

\begin{document}

\title{Calibration of  beam position monitors for high energy
  accelerators based on average trajectories}



\author{Javier  Fernando Cardona}
\email[]{jfcardona@unal.edu.co}
\affiliation{Universidad Nacional de Colombia, Bogot\'a, Colombia} 


\date{\today}

\begin{abstract}
This article presents a method that uses turn-by-turn beam position
data and \textit{k}-modulation data  to measure the calibration factors of
beam position monitors in high energy accelerators. In this method,
new algorithms have been developed to reduce the effect of coupling
and other sources of uncertainty, allowing accurate estimates of the
calibration factors. Simulations with known sources of errors indicate
that  calibration factors can be recovered with an accuracy of 0.7\% rms  for
arc beam position monitors and an accuracy of 0.4\% rms for interaction region beam
position monitors. The calibration factors are also obtained from 
LHC experimental data and  are used to evaluate the effect this
calibration has on  a quadrupole correction estimated  with the
action and phase jump method for a interaction region of the LHC.
\end{abstract}
\maketitle

\section{Introduction}\label{intro}
Beam position monitor (BPM) calibration is important for various
techniques  that measure optical parameters in accelerators, such as
quadrupole errors,  beta functions, and others. In this
paper a method to find those calibration factors, partially based on
the  tools used in action and phase jump (APJ) analysis,  is developed
for a high energy accelerator such as the LHC.
This method has three parts: the first part is used to find the
calibration factors of arc BPMs and the other two are used to find the
calibration factors of  high-luminosity interaction region (IR)
BPMs. The first part uses a  measured beam position  $z_{meas}$ and a true beam position  $z_{true}$
so that the calibration factors $C_i$  are found with
\begin{equation}
C_i = \frac{z_{meas}}{z_{true}},
\label{cal1}
\end{equation}
where $i$ is the BPM index, $z$ is the horizontal or vertical  component
of the beam position and  $z_{true}$  is estimated with
\begin{equation}
z = \sqrt{2 J_c \beta_r(s)} \sin\left[ \psi_r(s) -\delta_c \right],
\label{betatron}
\end{equation}
where $\beta_r(s)$ and  $\psi_r(s)$ are the lattice functions with
all gradient errors included,  and $J_c$ and $\delta_c$ are the action and phase
constants. Electronic noise, uncertainties in the determination of the
lattice functions, and  BPM calibration factors have been identified as the
main sources of uncertainty in Eq.~(\ref{betatron}). Although is not
possible to completely suppress the effect of these sources of
uncertainty, significant reductions can be achieved by using
average trajectories~\cite{jfc_prab17}, the most up-to-date techniques for finding
lattice function~\cite{softwarelhc2019}, and  statistical techniques as described in Sec.~IV of
reference~\cite{cardona_arxiv2020}. Several other improvements that
can be made when using Eq.~(\ref{betatron}) are studied in this paper. For
example, the sensitivity to uncertainties in $\psi_r(s)$ and
$\delta_c$ can be almost  completely suppressed using multiple average
trajectories, as explained in Sec.~\ref{phase_uncert}.  Coupling can
also affect the  validity of Eq.~(\ref{betatron}). This effect
is studied in Sec.~\ref{coup_eff} and compared with the other know
sources of uncertainty. Then, in Sec.~\ref{coup_red} is  shown how to
build average trajectories to significantly reduce the coupling
effects. All these improvements are used in simulations for which
arc  BPM gain errors  are intentionally introduced and then measured  to determine the accuracy of method in
Sec.~\ref{arc_bpms}. This section also presents  estimates of 
calibration factors from experimental  data and the effects 
this calibration has  on action and phase plots.

The second  and third parts of the method are introduced in Sec.~\ref{IR_bpms}
and, as in the previous case, accuracy studies  are carried out using
simulations. In addition, in this section, a list of calibration factors for
the IR1 BPMs is obtained from experimental data. Finally, as an
application of the presented calibration method, the sensitivity of APJ  to BPM
calibrations  is evaluated in Sec.~\ref{quad_corrs}. 
\section{Reducing the effect of  $\psi_r(s)$ and $\delta_c$
  uncertainties}\label{phase_uncert}
 Equation~(\ref{betatron}) may  be susceptible to $\delta_c$ uncertainties. This dependency can be minimized if $\delta_c$ is chosen
such that 
\begin{equation}
  \psi_r(s) -\delta_c  = p \frac{\pi}{2},
\label{sin1}
\end{equation}
where $p$ is an odd, positive or negative number. A particular average trajectory will not meet this condition for
all BPMs in the ring since $\delta_c$ is constant. however, it is possible to build an average trajectory for
every BPM in the ring  such that the condition~(\ref{sin1}) can always be
met. This procedure involves the construction of  several hundred average
trajectories, which can be time-consuming and resource-intensive.  Instead, some average trajectories can be built with equally spaced
$\delta_c$ values, and the average trajectory for which $\delta_c$ is
closest to meeting condition~(\ref{sin1}) is chosen as the optimal
trajectory for a particular BPM. Simulations indicate that an average
trajectory with $\delta_c$ 15 degrees apart from the condition~(\ref
{sin1}) is still good enough to hide any possible dependence of the
Eq.~(\ref{betatron}) on the  uncertainties of $\delta_c$. This means that only 24 average trajectories are needed. In practice,
several average trajectories out of 24 are chosen to estimate the
calibration constant for a particular BPM. The criterion for selecting these trajectories is
\begin{equation}
\left| \sin\left[ \psi_r(s) -\delta_c \right] \right| < 0.9,
\end{equation}
which still provides enough independence from  the
uncertainties of $\delta_c$. It should also be noted that following this
procedure, the propagated uncertainty in  Eq.~(\ref{betatron})  due to
the uncertainties of $\psi_r(s)$ also become negligible. 
\section{Coupling and the action and phase constants}\label{coup_eff}
\begin{figure}[h]
\centering
\includegraphics{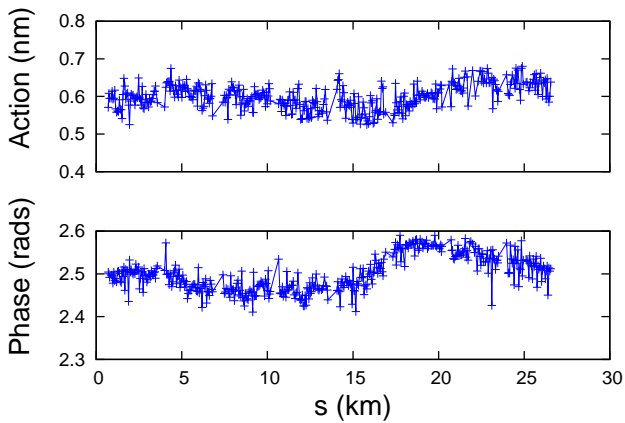}
\caption{\label{apexp} Action and phase plots of an average trajectory obtained
  from an experimental TBT data set  collected at the 2016 LHC run.}
\end{figure}

\begin{figure}[h]
\centering
\includegraphics{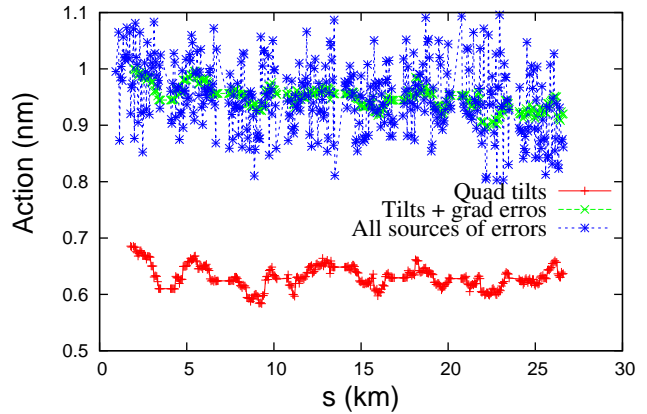}
\caption{\label{apsimu}  Action plots of simulated average trajectories
with coupling and other sources of errors.}
\end{figure}

Action and phase as a function of the axial coordinate $s$ are
expected to be  horizontal straight 
lines with values equal to $J_c$ and $\delta_c$. However, action and
phase plots obtained with 2016 LHC turn-by-turn (TBT) data
and lattice functions measured with the most up-to-date techniques~\cite{softwarelhc2019}
show small variations, as can be seen in Fig.~\ref{apexp}. 
These variations are a combination of slow and fast oscillations that can be
understood by simulations with different sources of errors. The
slow oscillations can be attributed to quadrupole tilt errors, as can
be seen from the red curve of Fig.~\ref{apsimu}. This curve corresponds
to the action plot of a  simulated  average trajectory
generated  with a  quadrupole tilt error distribution  with 2 mrad
standard deviation. 
The fast oscillations can be attributed to   BPM gain
and  noise errors and  uncertainties related to the determination of
the lattice functions. This is confirmed by the action plot (blue curve of Fig.~\ref{apsimu}) of a
simulated TBT data set generated with error distributions  with the currently
accepted rms values for the LHC: 3\% rms BPM gain
errors~\cite{bpmcal2020},  0.1 mm rms  BPM noise~\cite{malina_noise},
1\% rms uncertainties in the determination of the beta
functions~\cite{al_prst15,analyNBPM},  6 mrads rms uncertainties in
the determination of the betatron phases~\cite{psk_ipac16}, 2 mrad rms
quadrupole tilt errors, and $5*10^{-6}$ $m^{-2}$ rms gradient errors. It should
be noted that the amplitude of the slow oscillations can be comparable
to the amplitude of the fast oscillations, indicating that
the quadrupole tilt errors may be as important as the other sources of  errors
in accurately finding $J_c$ and $\delta_c$. It should also be noted
that the amplitude of the fast oscillations in the simulations is
larger than in the experimental data. This may  indicate that one
or all of the sources of these oscillations are smaller than the
currently accepted values. In Sec.~\ref{arc_bpms}, in fact, it is
found that the rms values of the calibration factors are somewhat smaller than
the 3\% mentioned earlier. 

Gradient errors alone shift the action and
phase plots vertically, changing the $J_c$ and $\delta_c$ values that can be
estimated from these plots. These changes, however, do not affect the
estimate of $z_{true}$ as long as the lattice functions that include
the gradient errors are used in Eq.~(\ref{betatron}).  The
displacement of the action plot can be seen by  comparing the red
and the green curves in Fig.~\ref{apsimu}. The green curve
is an action plot obtained with the 2 mrad rms
quadrupole tilt error distribution used in  the red curve plus a $5*10^{-6}$
$m^{-2}$ rms gradient error distribution. 
\section{ Building average trajectories  to reduce the effect of
  coupling}\label{coup_red}
\begin{figure}[h]
\centering
\includegraphics{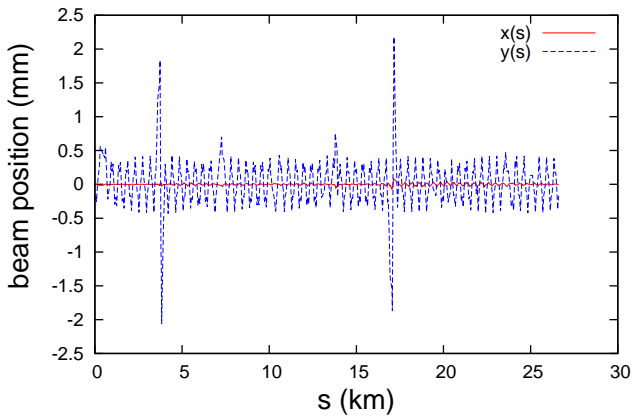}
\caption{\label{trajexp_max1plane}  Average trajectory obtained by
  selecting one-turn trajectories from experimental TBT data that satisfy the condition~(\ref{condi}) only
  in the vertical plane. As a consequence,
  the oscillations in the vertical plane are significantly larger than in
  the horizontal plane. It is also possible to built average
  trajectories with large oscillations in the horizontal plane and
  significantly smaller in the vertical plane.}
\end{figure}

\begin{figure}[h]
\centering
\includegraphics{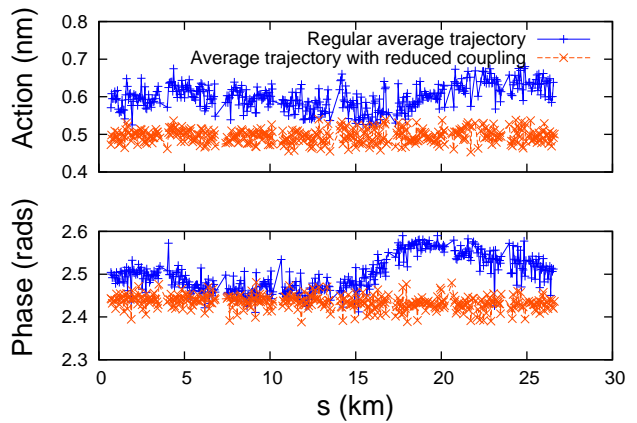}
\caption{\label{apexp_comp} Action and phase plots obtained with average
  trajectories built by applying condition \ref{condi} to both planes and
  to a single plane.  The slow oscillations, corresponding  to quadrupole 
 tilt errors, are significantly decreased with the new average trajectories.}
\end{figure}

Average trajectories are built by selecting trajectories  from a
TBT data set according to (complete procedure in Sec.~V of~\cite{jfc_prab17})
\begin{equation}
\left | \tilde{\delta}_z(n_m) -\tilde{\delta}_z(n) \right | <
\frac{\pi}{2}, 
\label{condi}
\end{equation}
where  $\tilde{\delta}_z(n)$ is the phase (as defined in~\cite{jfc_prab17}) associated with the trajectory
with turn number $n$,  and
\begin{equation}
\tilde{\delta}_z(n_m) =\psi_z(s_e) -p\frac{\pi}{2},
\label{defi}
\end{equation}
where $\psi_z(s_e)$ is the nominal betatron phase at the axial position
where the  average trajectory should be a maximum, and $p$  is an
odd, positive or negative number.  Regular average trajectories are
built using one-turn trajectories that satisfy the
condition~(\ref{condi})  in both planes  simultaneously. As a
consequence,  about a thousand of  one-turn trajectories are selected from the 6600
turns contained on a TBT data set. Now, if this condition is imposed
to only one plane, the number of  selected trajectories increases to
half the  total number of  trajectories in the TBT data set.  More
importantly, the average trajectory in the plane for which the
condition is not imposed tends to be negligible.  If this procedure is applied to an experimental TBT data set (the same one used to obtain Fig.~\ref{apexp}), the corresponding
average trajectory has significantly smaller oscillations in
one plane than in the other, as expected (Fig.~\ref{trajexp_max1plane}).
Significantly reducing the amplitude of the oscillations in one of the
planes also reduces the effect of linear coupling in the other plane,
as can be seen in the action and phase plots in Fig. ~\ref {apexp_comp}.
In addition to the average trajectory with a maximum at $s_e$ (max trajectory), it is also
possible to built an average trajectory with a minimum at   $s_e$ (min
trajectory). These two trajectories can be subtracted to obtain a max
trajectory that now is built from all the 6600 turns in the TBT data
set. 

\section{Measuring Gain Errors in  Arc BPMs}\label{arc_bpms}
All improvements mentioned in previous sections are used to find arc
BPM calibration factors with Eqs.~(\ref{cal1})~and~(\ref{betatron}), where all relevant
variables  are found from average trajectories derived from TBT
data and lattice functions. Both,  simulations and experimental
analysis are presented in the following subsections where two
conventions are adopted: first, the ``measured''
calibration factors  correspond to the factors obtained with
Eqs.~(\ref{cal1})~and~(\ref{betatron}) regardless of whether simulated
or experimental data is used, and second, the gain errors
$\varepsilon_{g,i}$ and the calibration factors $C_i$ are related by
\begin{equation}
\varepsilon_{g,i} = C_i -1
\end{equation} 
\subsection{Arc BPM calibration factors from simulations}
\begin{figure}[h]
\centering
\includegraphics{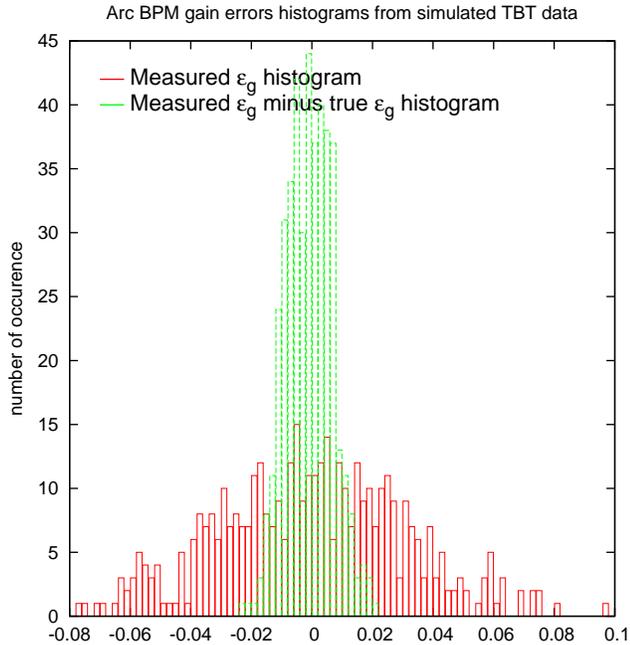}
\caption{\label{histo_arc_bpm_erry} The red  histogram corresponds to
  the  measured gain errors of the arc BPMs obtained from a simulated TBT data
  set. The standard deviation of the distribution is around 3\%  as
  expected. The green  histogram is the difference between   the
  measured BPM gain errors and the true gain errors used in the simulation. The
  standard deviation of this histogram indicates that the accuracy of
  the calibration for the arc BPMs is  approximately  0.8\% rms.}
\end{figure}

To evaluate the accuracy  (as
defined in~\cite{metro}\footnote{closeness of the agreement between
  the result of a measurement and a true value of the measurand}) of this part of the calibration method,
simulated TBT data  with the errors listed in Table~\ref{errs_simu}
are generated with MADX~\cite{mad}. The new lattice functions (nominal lattice
plus gradient errors) are also generated by MADX and, in addition, the
uncertainties associated with the determination of the lattice functions listed in
Table~\ref{errs_simu} are added. TBT data and lattice functions are
then used to obtain the action and phase plots and the measured BPM
gain errors.
\begin{table}[h]
\caption{\label{errs_simu} Rms values of known errors in the LHC
  lattice (first three rows) and uncertainties associated with the
  determination of lattice functions (last two rows).}
\begin{ruledtabular}
\begin{tabular}{l l c}
\multicolumn{1}{c}{Errors}
 &
\multicolumn{1}{c}{Rms value}
&
\multicolumn{1}{c}{Extracted from:}\\
\colrule
Gradients & $5*10^{-6}$ $m^{-2}$  & \cite{ml_fol_19}  \\
BPM gains &  3\% & \cite{bpmcal2020}  \\
Arc BPMs noise &  0.1 mm & \cite{malina_noise}  \\
$\qquad \beta_r(s)$ & 1\% & \cite{al_prst15,analyNBPM}  \\
$\qquad \psi_r(s)$ &6 mrads & \cite{psk_ipac16}  \\

\end{tabular}
\end{ruledtabular}
\end{table}
An histogram of all the measured  arc BPM gain errors obtained in this simulation for the vertical plane of beam 2  can be seen
in~Fig.~\ref{histo_arc_bpm_erry} (red bars). The same figure also
shows the histogram  of the differences between the measured BPM gain
errors and their corresponding true values (green bars). These
histograms illustrate that calibration factors with an original 3\% rms
distribution can be reduced to a 0.8\% rms calibration factors distribution.

\begin{figure}[h]
\centering
\includegraphics{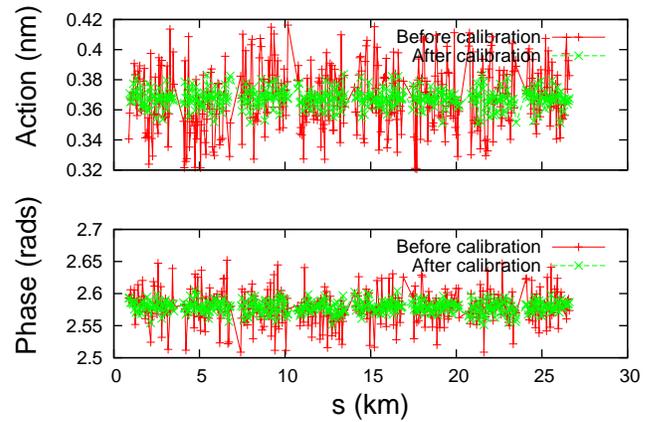}
\caption{\label{apsimu_calib} The action and phase plots of the
  simulated average trajectory with errors in Table~\ref{errs_simu} show
  a significant reduction of oscillations after calibration.}
\end{figure}

If the measured gain errors  are used to calibrate the original
simulated TBT data set, clear reductions in the variations of their
corresponding action and phase plots can be seen ( Fig.~\ref{apsimu_calib} ). 
\subsection{Arc BPM calibration factors from experimental data}
\begin{figure}[h]
\centering
\includegraphics{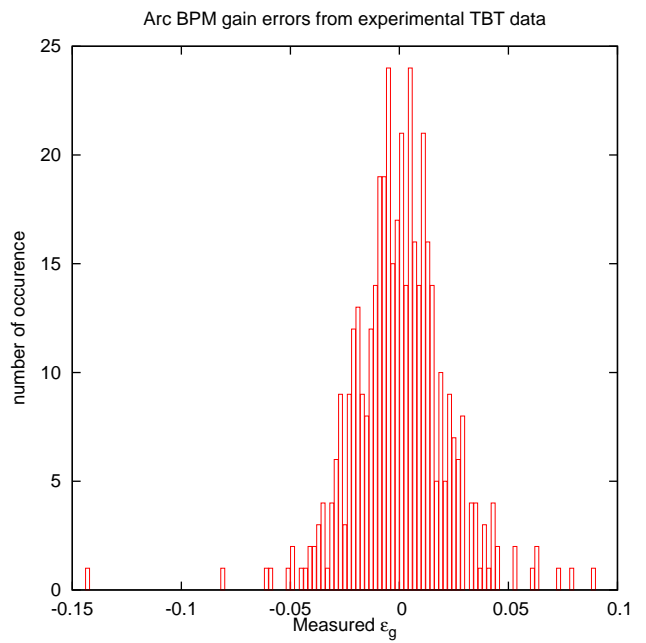}
\caption{\label{histoexp_arcbpms} Histogram of arc BPM gain errors
  (vertical plane) measured from an experimental TBT data of beam
  2. Calibration factors  in this histogram are distributed within a
   standard deviation of 2.2\% rms (2.3\% rms in the horizontal
  plane).}
\end{figure}

\begin{figure}[h]
\centering
\includegraphics{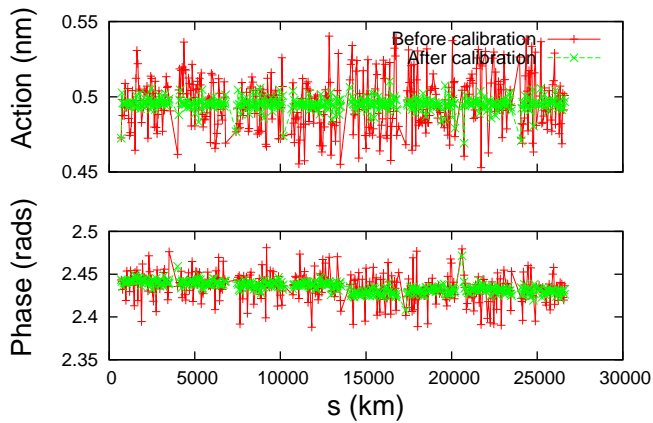}
\caption{\label{apexp_calib} Action and phase plots of an experimental
  TBT data set before and after the calibration factors in Fig.~\ref{histoexp_arcbpms} are
  applied.}
\end{figure}

A few TBT data sets taken during the 2016 LHC run are used to find the
calibration factors of the arc BPMs. These TBT data sets were taken
after global and local coupling corrections were applied on the IRs, but there
were no quadrupole corrections for gradient errors in the IRs. The
lattice functions are obtained directly from the same TBT data sets
using the most up-to-date algorithms, currently used in the LHC and
automatically  provided by the  orbit and measurement correction
(OMC) software~\cite{softwarelhc2019}.  Once the experimental TBT data and lattice
functions are available, same procedure used to obtain calibration
factors from simulated data is also used with  5 TBT data set of
beam 1 and 5 TBT data sets of beam 2. As an example, the gain error
histogram for the BPMs in the vertical plane of beam 2 is shown in Fig.~\ref{histoexp_arcbpms}.   
This histogram indicates that the rms gain error
in the arc BPMs is around 2\%, which is slightly smaller than the 3\%
reported in~\cite{bpmcal2020}.  Using the measured gain errors, the
experimental TBT data set is calibrated, leading to cleaner action and
phase plots, as seen in Fig~\ref{apexp_calib}.
\section{Measuring Gain Errors in IR BPMs}\label{IR_bpms}
In each  quadrupole triplet of the low $\beta^*$ IRs there are 3 BPMs:
BPMSW, BPMS and BPMYS.  To find their calibration factors two
different methods are used.
\subsection{Method to find calibration factors of BPMSWs}
The calibration method for these BPMs is essentially the same as for
the BPM arcs. The difference is that the beta functions are
estimated from \textit{k}-modulation experiments~\cite{kmod}. These experiments provide the  minimum value of the beta function 
between the triplets $\beta_w$ (commonly known as the beta function at the waist), and  the distance between the position of the waist and  the center of the
inter-triplet space $w$ (commonly known as the waist shift). The beta
functions at the BPMSWs are 
\begin{eqnarray}
\beta(s^{\pm}_b) = \beta_w + \frac{(L \pm w)^2}{\beta_w}
\label{betaw}
\end{eqnarray}
where  $s^{\pm}_b$ are the axial position of the BPMSWs located in the
left and right triplets and $L$ is half the length of the inter-triplet
space. The equation~(\ref{betaw}) leads to
very accurate measurement of the $\beta$ functions at the two BPMSWs,
which  should also allow a more accurate BPM calibration. 

\subsection{Method to find calibration factors of  BPMS and BPMYS}
For  BPMS and BPMYS, the \textit{k}-modulation technique is currently not
available, but a  modification of the  method for finding the calibration factors can
take advantage of the  accurate calibration of the  BPMSWs.

Suppose that a particle of beam 1,  coming from the inter-triplet space
of IR1, passes through  BPMSW registering a beam position $z(s_b)$. The particle
then passes through the first quadrupole of the right triplet 
(Q1) and then arrives to BPMS, where a beam position $z(s_s)$ is
recorded.   According to the action and phase method, any one-turn particle trajectory can be  described by 
\begin{equation}
z(s) = \sqrt{J(s) \beta_n(s)}\sin \left[ \psi_n(s) -\delta(s) \right], 
\label{eqapj}
\end{equation}
where the subscripts $n$ are used to refer to the nominal
variables. Hence,
\begin{eqnarray}
\label{apjbpms}
z(s_b) & = & \sqrt{J(s_b) \beta_n(s_b)} \sin \left[ \psi_n(s_b) -
\delta(s_b) \right]\\
\label{apjbpms2}
z(s_s) & = & \sqrt{J(s_s) \beta_n(s_s)} \sin \left[ \psi_n(s_s) -
\delta(s_s) \right].
\end{eqnarray}
On the other hand, $z(s_s)$ can also be expressed based on the
original action and phase $J(s_b)$ and $\delta(s_b)$ plus the
kick $\theta$ experienced by the particle due to a magnetic error
present in Q1 at $s_e$ (see Eq.~(1) of reference~\cite{jfc_prab17})
 \begin{eqnarray}
\label{xplus}
z(s_s) &= &\sqrt{J(s_b) \beta_n(s_s)} \sin \left[ \psi_n(s_s) -
  \delta(s_b) \right]+\\
 & &\theta \sqrt{\beta_n(s_s) \beta_n(s_e)} \sin
 \left[ \psi_n(s_s)-\psi_n(s_e) \right], \nonumber
\end{eqnarray}
 The phase advance between $s_s$ and $s_e$ is negligible
and hence
\begin{equation}
z(s_s) = \sqrt{J(s_b) \beta_n(s_s)} \sin \left[\psi_n(s_s) -
  \delta(s_b)\right].
\label{xapj2}
\end{equation}
Finally, using Eqs.~(\ref{apjbpms}) and~(\ref{xapj2})
\begin{equation}
z(s_s) = z(s_b) \sqrt{\frac{\beta_n(s_s)}{\beta_n(s_b)}}\frac{\sin
  \left[\psi_n(s_s) -\delta(s_b)\right]}{\sin \left[\psi_n(s_b) -\delta(s_b)\right]},
\label{xss}
\end{equation}
which allows estimating $z(s_s)$ from the nominal lattice functions
and $z(s_b)$ that is already calibrated. $\delta(s_b)$ corresponds
to the phase in the inter-triplet space $\delta_t$ and   can be
estimated with the formulas developed and tested in~\cite{cardona_arxiv2020}.  A similar procedure can be used to
estimate the   beam position at BPMSY.  
\subsection{ IR BPM calibration factors from simulations}
\begin{figure}[h]
\centering
\includegraphics{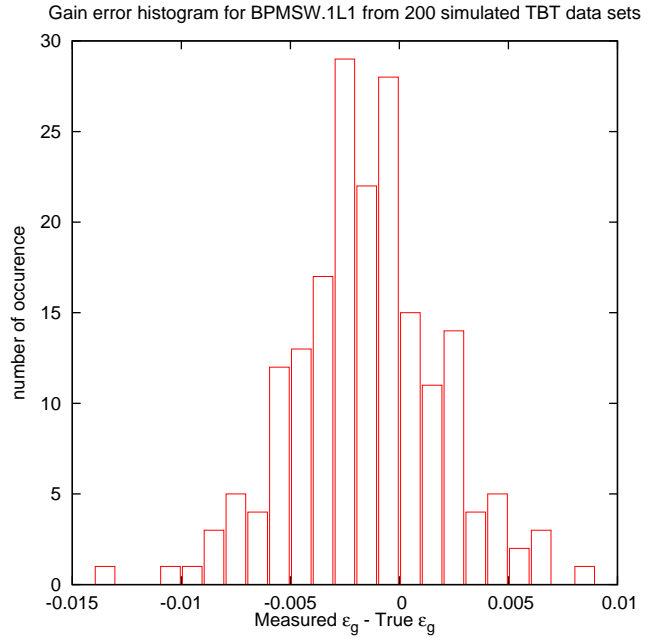}
\caption{\label{histoIRbmps} Histogram of the differences
  between the measured gain errors and the true gain errors in BPMSW.1L1
  for 200 hundred simulations with the random errors in Table~\ref{errs_sim2} . The
  standard deviation of this histogram indicates that the  calibration
  accuracy   for BPMSW.1L1 is approximately 0.3\% rms with the method
  presented in this section. }
\end{figure}

Two hundred simulated TBT data  sets with the errors listed in Table~\ref{errs_sim2} are generated with MADX  to
asses the accuracy of the calibration methods presented in this section. 
\begin{table}[h]
\caption{\label{errs_sim2} In addition to the  errors and uncertainties listed in Table~\ref{errs_simu},
the uncertainties associated to k-modulation experiments are included
since the beta functions derived from these experiments are used  for
calibration of IR BPMs.}
\begin{ruledtabular}
\begin{tabular}{l l  c}
\multicolumn{1}{c}{Errors}
 &
\multicolumn{1}{c}{Rms value}
&
\multicolumn{1}{c}{Extracted from:}\\
\colrule
Grads & $5*10^{-6}$ $m^{-2}$  &  \cite{ml_fol_19} \\
BPM gains &  3\% & \cite{bpmcal2020}  \\
Arc BPMs noise &  0.1 mm & \cite{malina_noise}  \\
$\qquad \beta_r(s)$ & 1\% & \cite{al_prst15,analyNBPM}  \\
$\qquad \psi_r(s)$ &6 mrads & \cite{psk_ipac16}  \\
Trip. quad grads & $2*10^{-5}$ $m^{-2}$  &  \cite{lhc_2015} \\
Match quad grads & $1*10^{-4}$ $m^{-2}$  & \cite{cardona_arxiv2020}  \\
$\qquad w_r$ & 1 cm &  \textit{k}-modulation experiments\\
$\qquad \beta_{w_r}$ & 0.3 mm &  \textit{k}-modulation experiments\\
\end{tabular}
\end{ruledtabular}
\end{table}
Random calibration factors with a standard deviation of 3\% rms are
assigned to the triplet BPMs plus a systematic shift of 5\% with
respect to the calibration factors of the arcs (as suggested by~\cite{bpmcal2020}). Since
there are two hundred simulations, there are two hundred measured calibration
factors obtained with Eq.~(\ref{xss}) and two hundred true calibration
factors for every IR BPM. The rms differences of these two quantities
are reported in  Table~\ref{acur_arcbpms1} for the  six  beam-2 BPMs
in IR1. Also, Fig.~\ref{histoIRbmps} shows a histogram of the measured
gain error minus the true gain errors for BPMSW.1L1. Similar histograms can be found for the
other 5 BPMs.
\begin{table}[h]
\caption{\label{acur_arcbpms1} Rms differences between measured calibration factors and  true calibration factors  for 200
    simulations. The errors listed in Table~\ref{errs_sim2} are added to
    each simulation according to a Gaussian distribution with the indicated standard deviations.}
\begin{ruledtabular}
\begin{tabular}{l c}
 &
\multicolumn{1}{c}{Rms accuracy}\\
BPM
&
\multicolumn{1}{c}{(\%)}\\
\colrule
BPMSW.1L1.B2   &   0.34 \\
BPMSW.1R1.B2  &   0.3  \\
BPMSY.4L1.B2  &    0.42 \\
BPMS.2L1.B2 &     0.23\\
BPMS.2R1.B2   &   0.28\\
BPMSY.4R1.B2 & 0.28\\
\end{tabular}
\end{ruledtabular}
\end{table}

\subsection{IR BPM calibration factors from experimental data}
The same 5 experimental TBT data sets of beam 1 and the 5 TBT data sets of beam 2 mentioned in
Sec.~\ref{arc_bpms} are used to find the calibration factors for the IR
BPMs. Furthermore, data from \textit{k}-modulation experiments performed
simultaneously while taking the experimental TBT data are used to
estimate the beta functions in the BPMSWs with Eq.~(\ref{betaw}). These
analyses finally  lead to calibration factors for the 6 triplet BPMs
of beam 1 and the six  triplet BPMs of beam 2 in both planes, as can be
seen in Table~\ref{calf_b1}~\ref{calf_b2}.

\begin{table}[h]
\caption{\label{calf_b1} Calibration Factor for IR1 BPMs  obtained from
  2016 experimental TBT and \textit{k}-modulation data of beam
  1.}
\begin{ruledtabular}
\begin{tabular}{l D{,}{\pm}{-1} D{,}{\pm}{6.4} }
& \multicolumn{2}{c}{Beam 1 Calibration Factors}\\
BPM Name &\multicolumn{1}{c}{HOR}
& \multicolumn{1}{c}{VERT}\\
\colrule
BPMSW.1L1  &   0.968,0.001 &   0.942,0.001\\
BPMSW.1R1 & 0.961,0.001 & 0.906,0.001 \\
BPMS.2L1  & 0.954, 0.001& 0.926,0.001\\
BPMS.2R1  &0.981, 0.001& 0.941,0.002\\
BPMSY.4R1 &   0.928,0.001 &   0.939,0.001\\
\end{tabular}
\end{ruledtabular}
\end{table}

\begin{table}[h]
\caption{\label{calf_b2} Calibration Factor for IR1 BPMs  obtained from
  2016 experimental TBT  and \textit{k}-modulation data of beam 2.}
\begin{ruledtabular}
\begin{tabular}{l D{,}{\pm}{-1} D{,}{\pm}{6.4} }
& \multicolumn{2}{c}{Beam 2 Calibration Factors}\\
BPM Name &\multicolumn{1}{c}{HOR}
& \multicolumn{1}{c}{VERT}\\
\colrule
BPMSW.1L1  &   0.953,0.002 &   0.942,0.001\\
BPMSW.1R1 & 0.935,0.001 & 0.944,0.001 \\
BPMSY.4L1 &   0.951,0.001 &  0.948,0.002\\
BPMS.2L1  & 0.942, 0.001& 0.941,0.001\\
BPMS.2R1  &0.947, 0.002& 0.954,0.002\\
BPMSY.4R1 &   0.955,0.001 &   0.979,0.001\\
\end{tabular}
\end{ruledtabular}
\end{table}

IR BPM Calibration factor are shifted about 5\% as reported
in~\cite{bpmcal2020}. The experimental uncertainty is estimated as the standard deviation of
the five measurements available for every calibration factor and it is
remarkably small.   

\section{BPM gain errors and quadrupole corrections in the
  IRs}\label{quad_corrs}
\begin{figure}[h]
\centering
\includegraphics{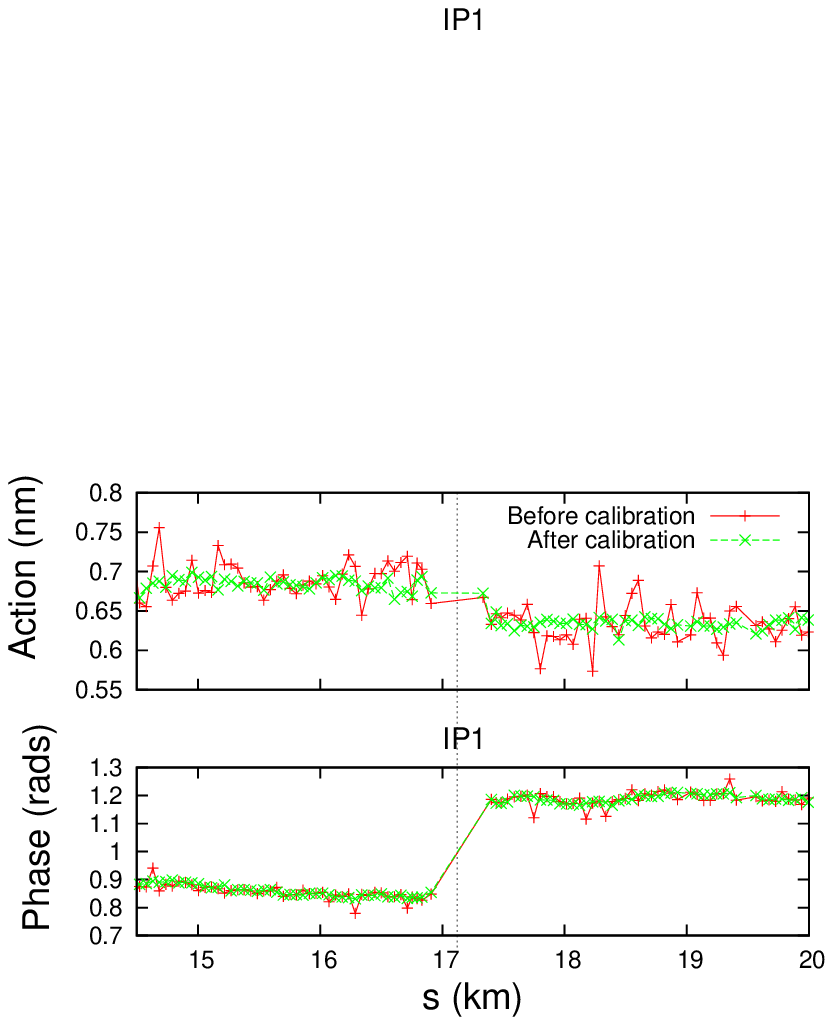}
\caption{\label{apj_IR1} Action and Phase jump in the horizontal plane
  of beam 2 at IR1  before and after applying the BPM calibration. The
  calibration procedure make it possible to define the jump more clearly, especially in the action plot.}
\end{figure}

Corrections to linear magnetic errors  in the IRs can be estimated
with the action and phase jumps that can be seen in action and phase
plots obtained with nominal lattice functions~\cite{jfc_prab17,
  cardona_arxiv2020}. Since these plots are derived from BPM
measurements, it is necessary to asses their sensitivity  to BPM
calibrations. To evaluate this sensitivity, the calibration factors
found for the arc  and IR BPMs in
Secs.~\ref{arc_bpms}~and~\ref{IR_bpms} are applied to the  same
experimental TBT data sets used in those sections and the corresponding action and phase plots are obtained
(Fig.~\ref{apj_IR1}).  Comparisons between the action and phase plots
before and after calibration show significant improvements,
particularly in the action plots.

\begin{figure}[h]
\centering
\includegraphics{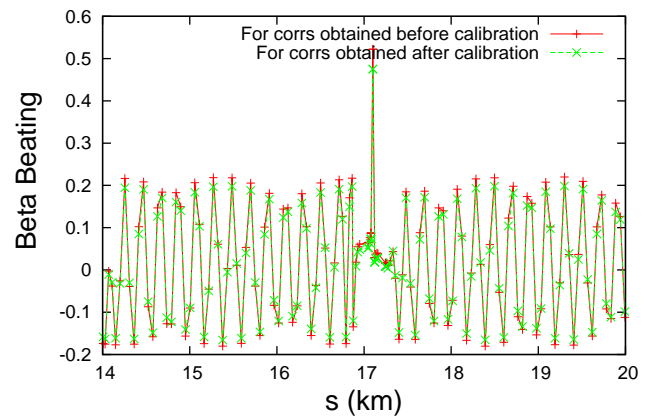}
\caption{\label{beat_corrs} Beta-beating generated by the corrections
  obtained before and after BPM calibration. Based on the small differences in the two plots, it can be stated that the two
  corrections are equivalent.}
\end{figure}

Also, since now the average trajectories are much larger in one plane than the other, the simplified expressions 
\begin{eqnarray}
{ B_1}_{x,e}  &= & - \frac{\theta_{x,e}}{x_e}, \\
{ B_1}_{y,e} &= & \frac{\theta_{y,e}}{y_e}
\end{eqnarray}
can be used to estimate the quadrupole components ${ B_1}_{z,e}$ of
the equivalent kick $\theta_{x,e}$ instead of Eqs.~(7) of~\cite{cardona_arxiv2020}.  Once these components are known, the
corrections are estimated  before and after calibration and  no
significant variations are found (Table~\ref{quad_corr}). The equivalence between
the two corrections can also be verified through the beta-beating that
they produce as can seen in Fig.~\ref{beat_corrs}.
\begin{table}[h]
\caption{\label{quad_corr} Quadrupole correction estimated for IR1 from experimental TBT data
  before and after the calibration procedure. Most of the proposed
  corrections in the 12 quadrupoles have only small variations between the
two cases.}
\begin{ruledtabular}
\begin{tabular}{c D{.}{.}{2.2} D{.}{.}{2.2} }
& \multicolumn{2}{c}{Correction strengths}\\
& \multicolumn{2}{c}{($10^{-5}$  $m^{-2}$)}\\
Magnet &\multicolumn{1}{c}{Before calibration}
& \multicolumn{1}{c}{After calibration}\\
\colrule
Q2L  &   1.24 &   1.19\\
Q2R &-0.84 &-0.76 \\
Q3L  &   1.44 &   1.36\\
Q3R  &-2.75&-2.60\\
Q4L.B1 &   11.3 & 10.2\\
Q4L.B2&-11.3 & 10.2\\
Q4R.B1 & -8.0 & -8.2\\
Q4R.B2 &   8.0 & 8.2\\
Q6L.B1 &-41.1  & -35.9 \\
Q6L.B2 &   34.2 &  29.9 \\
Q6R.B1 &   25.4  & 27.4 \\
Q6R.B2 &-22.2 & -24.0\\
\end{tabular}
\end{ruledtabular}
\end{table}

\section{Other Simulations}
The results of Sec.~\ref{arc_bpms} indicate that arc BPM gain errors are approximately
2.3\% rms. Also, the current number of turns has been increased to 10000,
which reduces the effect of electronic noise. Except fo these
changes, TBT data sets  are simulated with the same errors listed in
Table~\ref{errs_simu}. The calibrations factors of the BPM arcs can now be
recovered with 0.7\% accuracy instead of the original 0.8\% quoted in
Sec.~\ref{arc_bpms}. The IR BPMs calibrations also have a better associated
accuracy, as can be seen comparing Tables~\ref{accur_IRbpms}~and~\ref{acur_arcbpms1}.

\begin{table}[h]
\caption{\label{accur_IRbpms} Rms differences between measured calibration factors  and  true calibration factors  for 200
    simulations. Table~\ref{errs_sim2} is still used  in the simulations, except
    that gain errors are now 2.3\%  and  the electronic noise
    corresponds to what would exist for 10000 turns.}
\begin{ruledtabular}
\begin{tabular}{l c}
 &
\multicolumn{1}{c}{Rms accuracy}\\
BPM
&
\multicolumn{1}{c}{(\%)}\\
\colrule
BPMSW.1L1.B2   &   0.26 \\
BPMSW.1R1.B2  &   0.25  \\
BPMSY.4L1.B2  &    0.38 \\
BPMS.2L1.B2 &     0.19\\
BPMS.2R1.B2   &   0.26\\
BPMSY.4R1.B2 & 0.26\\
\end{tabular}
\end{ruledtabular}
\end{table}

\section{Conclusions}
A method  has been developed to find calibrations factors based on
average trajectories. Simulations show that the calibration factors for
arc BPMs can be recovered with an accuracy of 0.7\% rms and the calibatrion factors
for IR BPMs can be recovered with an accuracy of 0.4\% rms. The method has
been used to obtain the calibration factors of six BPMs of beam 1 and
six BPMs of beam 2 at the IR1 of the LHC. For these estimates, several
TBT data sets, measured lattice functions, and \textit{k}-modulation
measurements in the IRs are needed.

This  method has also been used to test the BPM calibration
sensitivity  of  the action and phase jump method. Although the
calibration helps to more clearly define the action and phase jump in the IR, its
effect on the estimation of corrections is negligible.

 \section*{Acknowledgments}
Many thanks to all members of  the optics measurement and correction team
(OMC) at CERN for support with their  \textit{k}-modulation  software, the
GetLLM  program, and  experimental data.

\bibliography{17}

\begin{thebibliography}{14}%
\makeatletter
\providecommand \@ifxundefined [1]{%
 \@ifx{#1\undefined}
}%
\providecommand \@ifnum [1]{%
 \ifnum #1\expandafter \@firstoftwo
 \else \expandafter \@secondoftwo
 \fi
}%
\providecommand \@ifx [1]{%
 \ifx #1\expandafter \@firstoftwo
 \else \expandafter \@secondoftwo
 \fi
}%
\providecommand \natexlab [1]{#1}%
\providecommand \enquote  [1]{``#1''}%
\providecommand \bibnamefont  [1]{#1}%
\providecommand \bibfnamefont [1]{#1}%
\providecommand \citenamefont [1]{#1}%
\providecommand \href@noop [0]{\@secondoftwo}%
\providecommand \href [0]{\begingroup \@sanitize@url \@href}%
\providecommand \@href[1]{\@@startlink{#1}\@@href}%
\providecommand \@@href[1]{\endgroup#1\@@endlink}%
\providecommand \@sanitize@url [0]{\catcode `\\12\catcode `\$12\catcode
  `\&12\catcode `\#12\catcode `\^12\catcode `\_12\catcode `\%12\relax}%
\providecommand \@@startlink[1]{}%
\providecommand \@@endlink[0]{}%
\providecommand \url  [0]{\begingroup\@sanitize@url \@url }%
\providecommand \@url [1]{\endgroup\@href {#1}{\urlprefix }}%
\providecommand \urlprefix  [0]{URL }%
\providecommand \Eprint [0]{\href }%
\providecommand \doibase [0]{http://dx.doi.org/}%
\providecommand \selectlanguage [0]{\@gobble}%
\providecommand \bibinfo  [0]{\@secondoftwo}%
\providecommand \bibfield  [0]{\@secondoftwo}%
\providecommand \translation [1]{[#1]}%
\providecommand \BibitemOpen [0]{}%
\providecommand \bibitemStop [0]{}%
\providecommand \bibitemNoStop [0]{.\EOS\space}%
\providecommand \EOS [0]{\spacefactor3000\relax}%
\providecommand \BibitemShut  [1]{\csname bibitem#1\endcsname}%
\let\auto@bib@innerbib\@empty
\bibitem [{\citenamefont {Cardona}\ \emph {et~al.}(2017)\citenamefont
  {Cardona}, \citenamefont {Garc\'{\i}a~Bonilla},\ and\ \citenamefont
  {Tom\'as~Garc\'{\i}a}}]{jfc_prab17}%
  \BibitemOpen
  \bibfield  {author} {\bibinfo {author} {\bibfnamefont {J.~F.}\ \bibnamefont
  {Cardona}}, \bibinfo {author} {\bibfnamefont {A.~C.}\ \bibnamefont
  {Garc\'{\i}a~Bonilla}}, \ and\ \bibinfo {author} {\bibfnamefont
  {R.}~\bibnamefont {Tom\'as~Garc\'{\i}a}},\ }\href {\doibase
  10.1103/PhysRevAccelBeams.20.111004} {\bibfield  {journal} {\bibinfo
  {journal} {Phys. Rev. Accel. Beams}\ }\textbf {\bibinfo {volume} {20}},\
  \bibinfo {pages} {111004} (\bibinfo {year} {2017})}\BibitemShut {NoStop}%
\bibitem [{\citenamefont {Carlier}\ \emph {et~al.}(2019)\citenamefont
  {Carlier}, \citenamefont {Coello}, \citenamefont {Dilly}, \citenamefont
  {Fol}, \citenamefont {Garcia-Tabares}, \citenamefont {Giovannozzi},
  \citenamefont {Hofer}, \citenamefont {Maclean}, \citenamefont {Malina},
  \citenamefont {Persson}, \citenamefont {Skowronski}, \citenamefont
  {Spitznagel}, \citenamefont {Tom\'as}, \citenamefont {Wegscheider},
  \citenamefont {Wenninger}, \citenamefont {Cardona},\ and\ \citenamefont
  {Rodriguez}}]{softwarelhc2019}%
  \BibitemOpen
  \bibfield  {author} {\bibinfo {author} {\bibfnamefont {F.}~\bibnamefont
  {Carlier}}, \bibinfo {author} {\bibfnamefont {J.}~\bibnamefont {Coello}},
  \bibinfo {author} {\bibfnamefont {J.}~\bibnamefont {Dilly}}, \bibinfo
  {author} {\bibfnamefont {E.}~\bibnamefont {Fol}}, \bibinfo {author}
  {\bibfnamefont {A.}~\bibnamefont {Garcia-Tabares}}, \bibinfo {author}
  {\bibfnamefont {M.}~\bibnamefont {Giovannozzi}}, \bibinfo {author}
  {\bibfnamefont {M.}~\bibnamefont {Hofer}}, \bibinfo {author} {\bibfnamefont
  {E.~H.}\ \bibnamefont {Maclean}}, \bibinfo {author} {\bibfnamefont
  {L.}~\bibnamefont {Malina}}, \bibinfo {author} {\bibfnamefont
  {T.}~\bibnamefont {Persson}}, \bibinfo {author} {\bibfnamefont
  {P.}~\bibnamefont {Skowronski}}, \bibinfo {author} {\bibfnamefont
  {M.}~\bibnamefont {Spitznagel}}, \bibinfo {author} {\bibfnamefont
  {R.}~\bibnamefont {Tom\'as}}, \bibinfo {author} {\bibfnamefont
  {A.}~\bibnamefont {Wegscheider}}, \bibinfo {author} {\bibfnamefont
  {J.}~\bibnamefont {Wenninger}}, \bibinfo {author} {\bibfnamefont
  {J.}~\bibnamefont {Cardona}}, \ and\ \bibinfo {author} {\bibfnamefont
  {Y.}~\bibnamefont {Rodriguez}},\ }in\ \href@noop {} {\emph {\bibinfo
  {booktitle} {Proceedings of the 2019 International Particle Accelerator
  Conference}}}\ (\bibinfo {year} {2019})\ p.\ \bibinfo {pages}
  {2773}\BibitemShut {NoStop}%
\bibitem [{\citenamefont {Cardona}\ \emph {et~al.}(2020)\citenamefont
  {Cardona}, \citenamefont {Rodríguez},\ and\ \citenamefont
  {Tomás}}]{cardona_arxiv2020}%
  \BibitemOpen
  \bibfield  {author} {\bibinfo {author} {\bibfnamefont {J.~F.}\ \bibnamefont
  {Cardona}}, \bibinfo {author} {\bibfnamefont {Y.}~\bibnamefont {Rodríguez}},
  \ and\ \bibinfo {author} {\bibfnamefont {R.}~\bibnamefont {Tomás}},\
  }\href@noop {} {\enquote {\bibinfo {title} {A twelve-quadrupole correction
  for the interaction regions of high-energy accelerators},}\ } (\bibinfo
  {year} {2020}),\ \Eprint {http://arxiv.org/abs/2002.05836} {arXiv:2002.05836
  [physics.acc-ph]} \BibitemShut {NoStop}%
\bibitem [{\citenamefont {Valdivieso}\ and\ \citenamefont
  {Tom\'as}(2020)}]{bpmcal2020}%
  \BibitemOpen
  \bibfield  {author} {\bibinfo {author} {\bibfnamefont {A.~G.-T.}\
  \bibnamefont {Valdivieso}}\ and\ \bibinfo {author} {\bibfnamefont
  {R.}~\bibnamefont {Tom\'as}},\ }\href {\doibase
  10.1103/PhysRevAccelBeams.23.042801} {\bibfield  {journal} {\bibinfo
  {journal} {Phys. Rev. Accel. Beams}\ }\textbf {\bibinfo {volume} {23}},\
  \bibinfo {pages} {042801} (\bibinfo {year} {2020})}\BibitemShut {NoStop}%
\bibitem [{\citenamefont {Malina}(2019)}]{malina_noise}%
  \BibitemOpen
  \bibfield  {author} {\bibinfo {author} {\bibfnamefont {L.}~\bibnamefont
  {Malina}},\ }\href@noop {} {\enquote {\bibinfo {title} {Lhc bpm performance:
  noise},}\ }\bibinfo {howpublished} {OMC-BI meeting} (\bibinfo {year}
  {2019})\BibitemShut {NoStop}%
\bibitem [{\citenamefont {Langner}\ and\ \citenamefont
  {Tom\'as}(2015)}]{al_prst15}%
  \BibitemOpen
  \bibfield  {author} {\bibinfo {author} {\bibfnamefont {A.}~\bibnamefont
  {Langner}}\ and\ \bibinfo {author} {\bibfnamefont {R.}~\bibnamefont
  {Tom\'as}},\ }\href {\doibase 10.1103/PhysRevSTAB.18.031002} {\bibfield
  {journal} {\bibinfo  {journal} {Phys. Rev. ST Accel. Beams}\ }\textbf
  {\bibinfo {volume} {18}},\ \bibinfo {pages} {031002} (\bibinfo {year}
  {2015})}\BibitemShut {NoStop}%
\bibitem [{\citenamefont {Wegscheider}\ \emph {et~al.}(2017)\citenamefont
  {Wegscheider}, \citenamefont {Langner}, \citenamefont {Tom\'as},\ and\
  \citenamefont {Franchi}}]{analyNBPM}%
  \BibitemOpen
  \bibfield  {author} {\bibinfo {author} {\bibfnamefont {A.}~\bibnamefont
  {Wegscheider}}, \bibinfo {author} {\bibfnamefont {A.}~\bibnamefont
  {Langner}}, \bibinfo {author} {\bibfnamefont {R.}~\bibnamefont {Tom\'as}}, \
  and\ \bibinfo {author} {\bibfnamefont {A.}~\bibnamefont {Franchi}},\ }\href
  {\doibase 10.1103/PhysRevAccelBeams.20.111002} {\bibfield  {journal}
  {\bibinfo  {journal} {Phys. Rev. Accel. Beams}\ }\textbf {\bibinfo {volume}
  {20}},\ \bibinfo {pages} {111002} (\bibinfo {year} {2017})}\BibitemShut
  {NoStop}%
\bibitem [{\citenamefont {Skowro\'nski}\ \emph {et~al.}(2016)\citenamefont
  {Skowro\'nski}, \citenamefont {Carlier}, \citenamefont {de~Portugal},
  \citenamefont {Garcia-Tabares}, \citenamefont {Langner}, \citenamefont
  {Maclean}, \citenamefont {Malina}, \citenamefont {McAteer}, \citenamefont
  {Persson}, \citenamefont {Salvant},\ and\ \citenamefont
  {Tom\'as}}]{psk_ipac16}%
  \BibitemOpen
  \bibfield  {author} {\bibinfo {author} {\bibfnamefont {P.}~\bibnamefont
  {Skowro\'nski}}, \bibinfo {author} {\bibfnamefont {F.}~\bibnamefont
  {Carlier}}, \bibinfo {author} {\bibfnamefont {J.~C.}\ \bibnamefont
  {de~Portugal}}, \bibinfo {author} {\bibfnamefont {A.}~\bibnamefont
  {Garcia-Tabares}}, \bibinfo {author} {\bibfnamefont {A.}~\bibnamefont
  {Langner}}, \bibinfo {author} {\bibfnamefont {E.}~\bibnamefont {Maclean}},
  \bibinfo {author} {\bibfnamefont {L.}~\bibnamefont {Malina}}, \bibinfo
  {author} {\bibfnamefont {M.}~\bibnamefont {McAteer}}, \bibinfo {author}
  {\bibfnamefont {T.}~\bibnamefont {Persson}}, \bibinfo {author} {\bibfnamefont
  {B.}~\bibnamefont {Salvant}}, \ and\ \bibinfo {author} {\bibfnamefont
  {R.}~\bibnamefont {Tom\'as}},\ }in\ \href@noop {} {\emph {\bibinfo
  {booktitle} {Proceedings of IPAC 2016}}}\ (\bibinfo {year}
  {2016})\BibitemShut {NoStop}%
\bibitem [{\citenamefont {of~the Joint Committee for Guides~in
  Metrology}(2008)}]{metro}%
  \BibitemOpen
  \bibfield  {author} {\bibinfo {author} {\bibfnamefont {W.~G.~.}\ \bibnamefont
  {of~the Joint Committee for Guides~in Metrology}},\ }\href@noop {} {\emph
  {\bibinfo {title} {Evaluation of measurement data — Guide to the expression
  of uncertainty in measurement}}},\ \bibinfo {organization} {BIPM, IEC, IFCC,
  ILAC, ISO, IUPAC, IUPAP and OIML} (\bibinfo {year} {2008})\BibitemShut
  {NoStop}%
\bibitem [{Note1()}]{Note1}%
  \BibitemOpen
  \bibinfo {note} {Closeness of the agreement between the result of a
  measurement and a true value of the measurand}\BibitemShut {NoStop}%
\bibitem [{\citenamefont {Grote}\ \emph {et~al.}(2015)\citenamefont {Grote},
  \citenamefont {Schmidt}, \citenamefont {Deniau},\ and\ \citenamefont
  {Roy}}]{mad}%
  \BibitemOpen
  \bibfield  {author} {\bibinfo {author} {\bibfnamefont {H.}~\bibnamefont
  {Grote}}, \bibinfo {author} {\bibfnamefont {F.}~\bibnamefont {Schmidt}},
  \bibinfo {author} {\bibfnamefont {L.}~\bibnamefont {Deniau}}, \ and\ \bibinfo
  {author} {\bibfnamefont {G.}~\bibnamefont {Roy}},\ }\href@noop {} {\emph
  {\bibinfo {title} {The MAD-X Program}}},\ \bibinfo {organization} {European
  Organization for Nuclear Research} (\bibinfo {year} {2015})\BibitemShut
  {NoStop}%
\bibitem [{\citenamefont {Fol}\ \emph {et~al.}(2019)\citenamefont {Fol},
  \citenamefont {de~Portugal}, \citenamefont {Franchetti},\ and\ \citenamefont
  {Tom\'as}}]{ml_fol_19}%
  \BibitemOpen
  \bibfield  {author} {\bibinfo {author} {\bibfnamefont {E.}~\bibnamefont
  {Fol}}, \bibinfo {author} {\bibfnamefont {J.~C.}\ \bibnamefont
  {de~Portugal}}, \bibinfo {author} {\bibfnamefont {G.}~\bibnamefont
  {Franchetti}}, \ and\ \bibinfo {author} {\bibfnamefont {R.}~\bibnamefont
  {Tom\'as}},\ }in\ \href@noop {} {\emph {\bibinfo {booktitle} {Proceedings of
  IPAC 2019}}}\ (\bibinfo {year} {2019})\ p.\ \bibinfo {pages}
  {3990}\BibitemShut {NoStop}%
\bibitem [{\citenamefont {Carlier}\ and\ \citenamefont {Tom\'as}(2017)}]{kmod}%
  \BibitemOpen
  \bibfield  {author} {\bibinfo {author} {\bibfnamefont {F.}~\bibnamefont
  {Carlier}}\ and\ \bibinfo {author} {\bibfnamefont {R.}~\bibnamefont
  {Tom\'as}},\ }\href {\doibase 10.1103/PhysRevAccelBeams.20.011005} {\bibfield
   {journal} {\bibinfo  {journal} {Phys. Rev. Accel. Beams}\ }\textbf {\bibinfo
  {volume} {20}},\ \bibinfo {pages} {011005} (\bibinfo {year}
  {2017})}\BibitemShut {NoStop}%
\bibitem [{\citenamefont {Persson}\ \emph {et~al.}(2017)\citenamefont
  {Persson}, \citenamefont {Carlier}, \citenamefont {Coello~de Portugal},
  \citenamefont {Garcia-Tabares~Valdiveso}, \citenamefont {Langner},
  \citenamefont {Maclean}, \citenamefont {Malina}, \citenamefont {Skowronski},
  \citenamefont {Salvant}, \citenamefont {Tom\'as},\ and\ \citenamefont
  {Garc\'ia~Bonilla}}]{lhc_2015}%
  \BibitemOpen
  \bibfield  {author} {\bibinfo {author} {\bibfnamefont {T.}~\bibnamefont
  {Persson}}, \bibinfo {author} {\bibfnamefont {F.}~\bibnamefont {Carlier}},
  \bibinfo {author} {\bibfnamefont {J.}~\bibnamefont {Coello~de Portugal}},
  \bibinfo {author} {\bibfnamefont {A.}~\bibnamefont
  {Garcia-Tabares~Valdiveso}}, \bibinfo {author} {\bibfnamefont
  {A.}~\bibnamefont {Langner}}, \bibinfo {author} {\bibfnamefont {E.~H.}\
  \bibnamefont {Maclean}}, \bibinfo {author} {\bibfnamefont {L.}~\bibnamefont
  {Malina}}, \bibinfo {author} {\bibfnamefont {P.}~\bibnamefont {Skowronski}},
  \bibinfo {author} {\bibfnamefont {B.}~\bibnamefont {Salvant}}, \bibinfo
  {author} {\bibfnamefont {R.}~\bibnamefont {Tom\'as}}, \ and\ \bibinfo
  {author} {\bibfnamefont {A.~C.}\ \bibnamefont {Garc\'ia~Bonilla}},\ }\href
  {\doibase 10.1103/PhysRevAccelBeams.20.061002} {\bibfield  {journal}
  {\bibinfo  {journal} {PRAB}\ }\textbf {\bibinfo {volume} {20}},\ \bibinfo
  {pages} {061002} (\bibinfo {year} {2017})}\BibitemShut {NoStop}%
\end{thebibliography}%
\end{document}